\documentclass[aps,prb,twocolumn,groupedaddress,showpacs]{revtex4}

\usepackage{graphicx}
\usepackage{dcolumn}
\usepackage{bm}

\begin{document}

\title{Large magneto-thermal effect and the spin-phonon coupling
in a parent insulating cuprate Pr$_{1.3}$La$_{0.7}$CuO$_4$}

\author{X. F. Sun}
\email[]{ko-xfsun@criepi.denken.or.jp}
\author{I. Tsukada}
\author{T. Suzuki}
\author{Seiki Komiya}
\author{Yoichi Ando}
\email[]{ando@criepi.denken.or.jp}
\affiliation{Central Research
Institute of Electric Power Industry, Komae, Tokyo 201-8511,
Japan.}

\date{\today}

\begin{abstract}

The magnetic-field ($H$) dependence of the thermal conductivity
$\kappa$ of Pr$_{1.3}$La$_{0.7}$CuO$_4$ is found to show a
pronounced minimum for in-plane fields at low temperature, which
is best attributed to the scattering of phonons by free spins that
are seen by a Schottky-type specific heat and a Curie-Weiss
susceptibility. Besides pointing to a strong spin-phonon coupling
in cuprates, the present result demonstrates that the
$H$-dependence of the phonon heat transport should not be naively
neglected when discussing the $\kappa(H)$ behavior of cuprates,
since the Schottky anomaly is ubiquitously found in cuprates at
any doping.

\end{abstract}


\maketitle

\section{Introduction}

To elucidate the nature of the superconducting ground state, the
quasiparticle (QP) transport properties are widely studied in
high-$T_c$ cuprates at low temperature ($T$), and the discussions
are often based on the magnetic-field ($H$) dependence of the
thermal conductivity $\kappa$.\cite{Krishana, Aubin, Ando1, Chiao,
Sun1, Ando2, Hawthorn} The most striking findings include a
``plateau" in the $\kappa(H)$ isotherm at low
temperature,\cite{Krishana, Aubin, Ando1} an increase in the QP
population due to the supercurrents around vortices,\cite{Chiao}
and the magnetic-field-induced QP localization in underdoped
cuprates.\cite{Sun1, Ando2, Hawthorn} A common assumption in these
previous works is that the phononic thermal conductivity
$\kappa_p$ is {\it independent} of $H$ and all the $H$-dependence
of $\kappa$ is of electronic origin. This assumption may seem
reasonable, because the vortex scattering of phonons (which is
important in conventional superconductors) somehow becomes
insignificant at low temperature in cuprates.\cite{Franz} However,
phonons could interact also with magnetic excitations and be
scattered, which would introduce some $H$-dependence in
$\kappa_p$. Therefore, it is important to elucidate how the
low-$T$ phonon heat transport is affected by magnetic excitations
in cuprates, particularly in the superconducting doping regime.
Remember, even though antiferromagnetic (AF) magnon excitations
are likely to be irrelevant in the superconducting regime,
Schottky anomalies in the specific heat, which are usually
attributed to excitations of localized free spins, are commonly
found in superconducting cuprates.\cite{Moler, Revaz, Wright,
Chen, Brugger, Hien, Schottky}

To gain insight into the possible $H$-dependence of $\kappa_p$, it
is useful to study parent insulating cuprates, where it is
expected that the electronic contribution is absent and $\kappa_p$
can be unambiguously studied; however, it has been discussed that
in AF insulating materials the magnon excitations can also act as
heat carriers or scatterers,\cite{Sun2, Hess, Hofmann, Takeya}
which brings complications to the analysis of the $\kappa(H)$ data
of insulating samples. In particular, Jin {\it et al.}\cite{Jin}
recently reported a large magnetic-field-induced enhancement of
$\kappa$ in Nd$_2$CuO$_4$ (NCO), a parent insulating cuprate in
the tetragonal $T'$-phase, and they proposed that the increase in
$\kappa$ gives evidence for the magnon heat transport becoming
active due to the closing of a magnon gap in the magnetic field.
Note that in NCO the magnons for the Cu spins should not
contribute to carrying heat because of a large ($\sim$5 meV) gap,
and thus it must be the Nd magnons that is relevant;\cite{Li} the
existence of the Nd-magnon contribution to the heat transport has
been backed up by a more recent study by Li {\it et al.}\cite{Li}
Considering these findings, to unambiguously investigate the $H$
dependence of $\kappa_p$, we turn to the
Pr$_{1.3}$La$_{0.7}$CuO$_4$ (PLCO) system where there is
essentially no magnetic contribution from the rare-earth ions.
Although it was found that a small magnetic moment is induced in
Pr$^{3+}$ ions (whose ground state is singlet and would be
effectively nonmagnetic),\cite{Sumarlin, Lavrov1} in
Pr$_{1.3}$La$_{0.7}$CuO$_4$ as much as 35\% of the Pr$^{3+}$ sites
are diluted by nonmagnetic La$^{3+}$ ions and thus the Pr magnons,
even if exist, should be strongly damped and give negligible
contribution to the heat transport. Also, the single crystals of
PLCO were found to show very clean phonon heat transport ({\it
e.g.}, $\kappa_p$ of PLCO is up to 5 times larger at 20 K compared
to La$_2$CuO$_4$,\cite{Sun3}) so that exotic scattering mechanisms
of phonons are expected to be rather easily seen in PLCO without
being masked by scatterings by disorder.

In this paper, we report detailed thermal conductivity, specific
heat, and magnetic susceptibility measurements of high-quality
Pr$_{1.3}$La$_{0.7}$CuO$_4$ (PLCO) single crystals. At low
temperature, the $\kappa(H)$ curve shows a striking dip feature,
which is different from previous data on similar $T'$-phase
compounds, Pr$_{2}$CuO$_4$ (PCO) and NCO.\cite{Jin,Sales} Our new
result on $\kappa(H)$ indicates that excitations of free spins,
evidenced by the specific heat and the magnetic susceptibility
data, are responsible for the peculiar $H$ dependence of the
phonon heat transport. Since the Schottky-type specific heat that
is indicative of the existence of free spins is found almost
ubiquitously in high-$T_c$ cuprates,\cite{Moler, Revaz, Wright,
Chen, Brugger, Hien, Schottky} the present result suggests that
the phonon heat transport can be $H$ dependent at low temperature
in a wide range of cuprate samples and one should be careful upon
discussing the QP properties based on $\kappa(H)$ behavior.
Physically, the present result points to a strong spin-phonon
coupling in cuprates, which might bear intriguing implications on
the mechanism of high-$T_c$ superconductivity.

\section{Experiments}

High-quality Pr$_{1.3}$La$_{0.7}$CuO$_4$ single crystals are grown
by the traveling-solvent floating-zone technique in flowing
oxygen.\cite{Sun3} The partial substitution of La for Pr not only
stabilizes the crystal growth\cite{Sun3, Fujita} but also disturbs
the possible Pr magnons and makes them irrelevant as heat
carriers, which is useful for the purpose of this work. For the
heat transport measurements, the crystals are cut into rectangular
platelets with a typical size of $2.5 \times 0.5 \times 0.1$
mm$^3$ (where the edges are made to be parallel to the
crystallographic axes within 1$^{\circ}$) and annealed in flowing
Ar to remove excess oxygen.\cite{Sun3}

The $a$- and $c$-axis thermal conductivities ($\kappa_a$ and
$\kappa_c$) are measured by a steady-state technique in a $^3$He
refrigerator.\cite{Sun1} Two RuO$_2$ chip sensors are used for the
``top" and ``bottom" thermometers on the sample. For the
magnetic-field dependent $\kappa(H)$ measurements, the base
temperature of heat sink is precisely controlled by an ac
resistance bridge (Linear Research LR-700) within 0.01\% accuracy
with another RuO$_2$ thermometer or a Cernox thermometer (for
above 10 K), and the magnetic-field-dependence of the thermometers
were carefully calibrated beforehand using a SrTiO$_3$ capacitance
sensor and a high-resolution capacitance bridge (Andeen Hagerling
2500A), which allow to keep the temperature unchanged with a very
high precision ($\pm$ 0.1 mK at 0.3 K and $\pm$ 1 mK above 1 K)
during the magnetic field sweeps. The relative uncertainty of the
$\kappa(H)$ measurement was confirmed to be typically less than
1\%. Specific heat and magnetic susceptibility measurements are
carried out using a Physical Properties Measurement System and a
SQUID magnetometer (Quantum Design), respectively.

\section{Results and discussions}

\subsection{Thermal conductivity}

\begin{figure}
\includegraphics[clip,width=7.5cm]{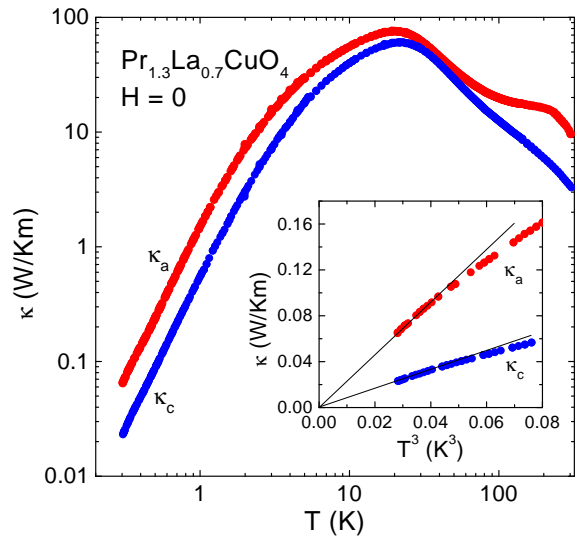}
\caption{Temperature dependences of $\kappa_a$ and $\kappa_c$ of
PLCO single crystals in zero field. Inset: $\kappa$ vs $T^3$ plot
of the low $T$ data below 0.43 K. Thin solid lines show the $T^3$
dependence expected in the boundary-scattering regime.}
\end{figure}

Figure 1 shows the temperature dependences of $\kappa_a$ and
$\kappa_c$ of PLCO in zero field down to 0.30 K, where the data
above 4 K are a reproduction from Ref. \onlinecite{Sun3}. As
already discussed, one can assume that both $\kappa_a$ and
$\kappa_c$ show purely phononic thermal conductivity in this
insulating compound, and the large magnitude of the phonon peak
evidences a high quality of our PLCO crystals.\cite{Sun3} The
inset of Fig. 1 shows that the boundary-scattering regime of
phonons (where $\kappa_p \sim T^3$) is achieved below $\sim$ 0.35
K.

\begin{figure}
\includegraphics[clip,width=8.5cm]{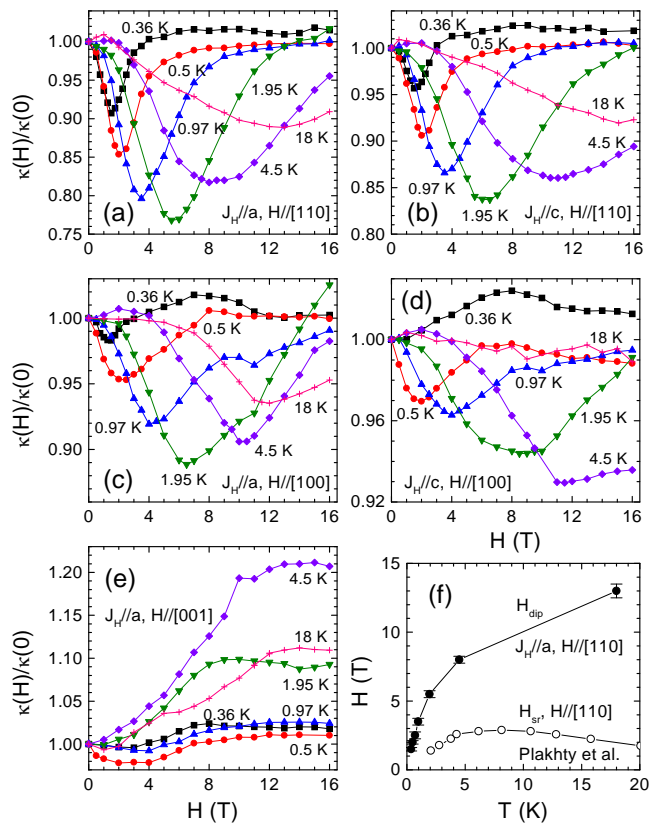}
\caption{(a)--(e) Magnetic field dependences of $\kappa$ in PLCO
single crystals for various configurations of the heat current
$J_H$ and the magnetic field. (f) Comparison of $H_{\rm dip}$
(where the $\kappa(H)$ curve shows a minimum) for $J_H \parallel
a$, $H \parallel$ [110] to the spin-reorientation field $H_{\rm
sr}$ for the same field direction (Plakhty {\it et al.}, Ref.
\onlinecite{Plakhty}).}
\end{figure}

Figure 2 shows our main result, the peculiar magnetic-field
dependences of $\kappa$ at temperatures from 0.36 to 18 K, for
several different configurations of heat current and magnetic
field. Notably, a pronounced dip in the $\kappa(H)/\kappa(0)$
curves is observed for in-plane fields at low tempratures. At
first sight, this may seem a typical $\kappa(H)$ behavior in a
spin-flop system where magnons act as phonon
scatterers:\cite{Buys} The zero-field $\kappa$ is purely phononic,
because there are no low-energy magnons due to the anisotropy gap
in the magnon dispersion; when the magnetic field is applied, the
Zeeman energy causes the magnon excitations to become gapless at
the spin-flop transition field $H_{\rm sf}$, but the gap opens
again at higher field; consequently, the magnon scattering of
phonons is the strongest at $H_{\rm sf}$ where the magnons are the
most populated, and causes a dip in the $\kappa(H)$
curve.\cite{Buys}

In the $T'$-phase compounds of NCO, PCO, and PLCO, the magnetic
structures are essentially identical and spin-flop (or
spin-reorientation) transition of the long-range-ordered Cu spins
has been studied by the neutron scattering:\cite{Sumarlin,
Lavrov1, Skanthakumar, Plakhty, Plakhty2, Petitgrand1,
Petitgrand2, Cherny} The Cu spins form a noncollinear structure
with the spin easy axes of [100] and [010] in zero-field;
depending on the magnetic-field direction, a first-order
transition (for $H \parallel$ [100]) or a continuous
spin-reorientation transition (for $H \parallel$ [110]) is
induced, and in both cases the high-field state has a collinear
spin structure. If one tries to interpret our $\kappa(H)$ data in
terms of this well-studied spin reorientation, several problems
become apparent: Most notably, the characteristic field for the
spin reorientation, $H_{\rm sr}$, is known to decrease upon
heating, while the ``dip" field, $H_{\rm dip}$, where the
$\kappa(H)$ curve shows a minimum, increases upon heating. We show
quantitative comparison between $H_{\rm sr}$ and $H_{\rm dip}$ for
$H \parallel$ [110] in Fig. 2(f), where it becomes clear that
$H_{\rm dip}$ is not likely to be related to the spin
reorientation. Also, $H_{\rm dip}$ is observed to be very close
for $H \parallel$ [100] and $H \parallel$ [110], while $H_{\rm
sr}$ is much larger for $H \parallel$ [100] than for $H
\parallel$ [110].\cite{Lavrov1, Plakhty2, Cherny} Furthermore,
the $H$-dependence of $\kappa_c$ is very similar to that of
$\kappa_a$, which is difficult to understand in the spin-flop
scenario because the $c$-axis phonons can hardly couple with the
magnons of the two-dimensional spin system. Therefore, the
peculiar $\kappa(H)$ behavior in PLCO is clearly {\it not} due to
the magnon scattering of phonons associated with the spin
reorientation.

It is known from previous neutron experiments\cite{Sumarlin,
Lavrov1} that the exchange field of the ordered Cu spins induces a
small ($\sim$ 0.1 $\mu_B$) ordered moment on Pr$^{3+}$ ions. One
may expect that the Pr magnons should contribute to the
$\kappa(H)$ behavior. Since the weak Pr moments are induced by the
interaction between Cu and Pr ions, the Pr magnons, if exist and
play a role (as either heat carriers or scatterers) in determining
the $\kappa(H)$ behavior, should affect $\kappa(H)$ in a way that
corresponds to the spin reorientation transition. Therefore, the
lack of correspondence between $H_{dip}$ and $H_{sr}$ means that
not only Cu magnons but also Pr magnons are irrelevant to the
observed $\kappa(H)$ behavior, which is likely due to the
considerable dilution of the Pr sites by nonmagnetic La$^{3+}$
ions.

\subsection{Specific heat}

\begin{figure}
\includegraphics[clip,width=8.5cm]{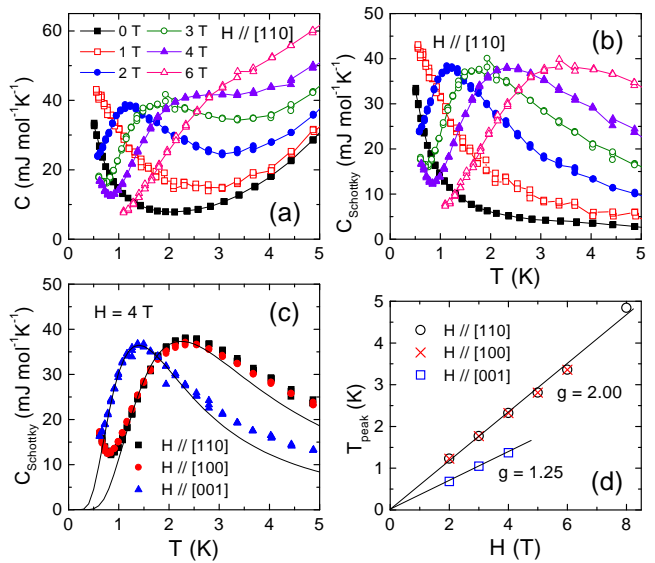}
\caption{(a) Temperature dependences of the specific heat of a
PLCO single crystal in various magnetic fields along [110]. (b)
Schottky anomaly for $H \parallel$ [110], obtained by subtracting
the phonon contribution from the raw data in panel (a). (c)
Schottky anomaly at 4 T for different orientations of $H$,
together with the theoretical curves expected for the Schottky
anomaly from 1\% of $s$ = 1/2 moments with the anisotropic
Land$\acute{e}$ factor $g$ = 2 (for [100] and [110]) and $g$ =
1.25 (for [001]). (d) Magnetic field dependences of the peak
temperature in the Schottky anomaly for the three directions.}
\end{figure}

To investigate the actual scatterers of phonons and the origin of
the peculiar $\kappa(H)$ behavior, the specific heat of PLCO is
measured at low temperature using a large single crystal ($\sim$
34 mg, Ar-annealed) from the same batch. As shown in Fig. 3, there
is a pronounced $H$-dependent enhancement, whose temperature and
field dependences are fully consistent with the two-level Schottky
anomaly due to free $s$ = 1/2 moments\cite{Schottky, Moler}
\begin{equation} C_{Schottky}(T,H) = n(\frac{g\mu_BH}{k_BT})^2
\frac{e^{g\mu_BH/k_BT}}{(1+e^{g\mu_BH/k_BT})^2}, \label{Schottky}
\end{equation}
where $g$ is the Land$\acute{e}$ factor. The concentration of free
spins is $n/R$, where $R$ is the universal gas constant. In our
data, the peak position of the Schottky anomaly, $T_{peak}$,
increases linearly with $H$ [see Fig. 3(d)], which precisely
follows Eq. (\ref{Schottky}) and allows us to determine the $g$
factor. Moreover, fitting of the data to Eq. (\ref{Schottky}) [see
Fig. 3(c)] gives the concentration of the free spins to be $\sim$
1\%, which is too large to be associated with impurities and
suggests the existence of some intrinsic mechanism to produce free
moments.

\subsection{Magnetic susceptibility}

\begin{figure}
\includegraphics[clip,width=8.5cm]{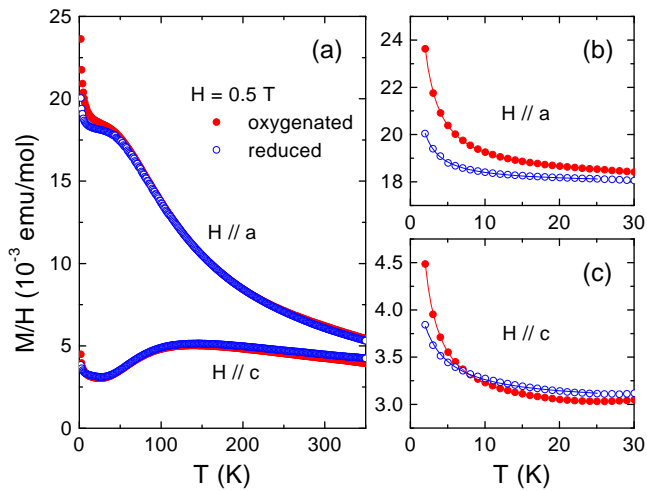}
\caption{(a) Temperature dependences of susceptibility $\chi_a$
and $\chi_c$ of a PLCO crystal in oxygenated (as-grown) and
reduced (Ar-annealed) states. Panels (b) and (c) magnify the
low-$T$ part; solid lines show fittings with the formula $\chi(T)
= \chi_0 + C/(T-\theta)$, where the first and second terms are
$T$-independent background and additional Curie-Weiss
contribution, respectively; the $C$ parameter for the reduced
sample is consistent with $\sim$ 1\% of $s$ = 1/2 free spins with
$g$ = 2 ($g$ = 1.25) for $H \parallel a$ ($\parallel c$).}
\end{figure}

Another evidence for the existence of free moments in PLCO is the
magnetic susceptibility ($\chi$) data. Figure 4 shows the
temperature dependences of $\chi$ for $H \parallel a$ and $H
\parallel c$ ($\chi_a$ and $\chi_c$) in 0.5 T field. Our data are
essentially consistent with those for PCO single
crystals,\cite{Hundley} including the low-$T$ upturn that is
observed in both $\chi_a$ and $\chi_c$. As theoretically discussed
by Aharony's group,\cite{Sachidanandam} the magnetic
susceptibility of R$_2$CuO$_4$ (R = Nd, Pr, or Sm) is governed by
the magnetic states of rare-earth ions. The ground multiplet of
Pr$^{3+}$ ion is $^3H_4$ in the tetragonal crystalline electric
field (CEF), so that Pr$^{3+}$ ions have a singlet ground state
with a well separated ($\sim$ 18 meV) first-excited
state.\cite{Sachidanandam} Therefore, the low-$T$ susceptibility
is predicted to be $T$-independent,\cite{Sachidanandam} and the
observed low-$T$ upturn, which is well described by the
Curie-Weiss law, must come from some free moments not considered
in the mean-field theory. Note that the sensitivity of the low-$T$
upturn in $\chi$ to the oxygen treatment, shown in Fig. 4,
suggests that the free moments do not originate from impurity
ions, but are produced by some mechanism inherent to PLCO.
Candidates for such a mechanism include an appearance of Pr$^{4+}$
ions in some particular crystal environment or some ``free"
Cu$^{2+}$ spins possibly produced in the AF domain
boundaries.\cite{Vaknin, Lavrov3} Incidentally, the analyses of
the Schottky anomaly and the Curie-Weiss contribution are
consistent with the same anisotropic $g$ factor, which gives
confidence that the two phenomena are of the same origin.

\subsection{Phonon scattering by paramagnetic moments}

Whatever the origin of the free moments, they are clearly
correlated with the $H$-dependence of $\kappa$ for in-plane
fields. Although there has been no report in the cuprate context,
it has been known that phonons can be scattered by free
paramagnetic moments that form a two-level system,\cite{Berman}
where the energy levels of the two states split in magnetic fields
by $\Delta E$ due to the Zeeman energy. In this situation, phonons
are typically scattered in the following way: A phonon with energy
$\Delta E$ excites the lower-level state of a magnetic moment to
the higher-level state and is absorbed by a direct process; then,
another phonon with the same energy (but with a different wave
vector) is subsequently emitted by the higher-level state as it
relaxes. Since the phonon spectrum has a (broad) maximum at $\sim
4k_{B}T$, this ``paramagnetic scattering" of phonons is the
strongest when the Zeeman energy $\Delta E$ is equal to $4k_{B}T$.
Naturally, this causes a dip at $H_{\rm dip}$ in the $\kappa(H)$
isotherm with $H_{\rm dip}$ roughly proportional to
$T$,\cite{Berman} which is actually the case in our data below 2 K
[see Fig. 2(f)]. Note that the free spins themselves do not carry
heat because there is no dispersive collective excitations due to
the negligibly small interactions between these diluted spins.

In passing, we note that the peculiar $\kappa(H)$ behavior in NCO
(Ref. \onlinecite{Jin}) may also be understood in terms of the
paramagnetic scattering of phonons without employing the magnon
heat transport, at least for $T >$ 1.5 K: It has been discussed
that the doublet ground state of the Nd$^{3+}$ ions in NCO is
split slightly ($\sim$ 3 K) in zero field due to the interaction
with Cu spins,\cite{Sachidanandam} which causes the Nd$^{3+}$ ions
to form a two-level system; hence, the phonons may well be
scattered by paramagnetic Nd$^{3+}$ ions at $T >$ 1.5 K, since the
Nd ions show Ne\'el order only below $\sim$ 1.5
K,\cite{Sachidanandam} and a sufficiently large $H$ can remove the
paramagnetic scatterings and enhance $\kappa$
significantly.\cite{Berman} Thus, although the magnons are likely
to be carrying heat below $\sim$ 1.5 K,\cite{Li} their role at
higher temperature had better be scrutinized.

We should point out that we do not currently have a good
understanding for the enhancement of $\kappa$ with $H
\parallel c$ shown in Fig. 2(e), but it may be that for this field
direction some additional heat carriers are induced and add a
major contribution on top of $\kappa_p(H)$. One possibility is a
remaining contribution of Pr magnons, although as much as 35\% of
the Pr$^{3+}$ sites are diluted by nonmagnetic La$^{3+}$ ions in
PLCO. Remember that La-free PCO single crystals had been reported
to exhibit a large enhancement of $\kappa$ with $H$,\cite{Sales}
which might share a common origin to the present behavior shown in
Fig. 2(e).

Although the interaction between phonons and the free spins has
been known for a long time,\cite{Berman} such phenomenon and its
impact on the heat transport have never been revealed for cuprate
materials. The present results demonstrate that the phonon heat
transport can be $H$ dependent at low temperature in high-$T_c$
cuprates whenever excitations of free spins (that cause a Schottky
anomaly in the specific heat) are present. The effect of the
paramagnetic scattering is particularly pronounced in PLCO,
probably because in this parent compound the phononic heat
transport is very clean and also the Schottky anomaly is large. In
the superconducting doping regime of cuprates, while the Schottky
anomaly is almost ubiquitously observed,\cite{Moler, Revaz,
Wright, Chen, Brugger, Hien, Schottky} its contribution is much
smaller than that in PLCO and also the phonons are additionally
scattered by doped carriers, both of which cause the contribution
of the paramagnetic scattering to be reduced. Nevertheless, it is
likely that some fraction of the $H$ dependence of $\kappa$ is due
to the paramagnetic scattering of phonons even in superconducting
samples, and therefore an analysis of $\kappa(H)$ should be done
very carefully. Note that, while {\it quantitative analyses} of
$\kappa(H)$ would be difficult in light of the paramagnetic
scattering of phonons, {\it qualitative comparisons} of the
$\kappa(H)$ behavior for different dopings \cite{Sun1,Ando2} would
still be legitimate, because the Schottky anomaly is observed
across the whole doping range without any clear doping dependence.

Lastly, the present result suggests, together with the curious
magnetic shape-memory effect in
La$_{2-x}$Sr$_{x}$CuO$_4$,\cite{Lavrov2} that the spin-phonon
coupling is rather strong in cuprates. The significance of the
spin-phonon coupling in {\it doped} cuprates, although still
remains to be verified, is rather naturally expected based on the
behavior of the parent compound. Because of this coupling, AF
fluctuations and phonons may couple in an unusual way in these
materials and this could be the reason why the mechanism of
high-$T_c$ superconductivity is so difficult to understand. In
this regard, it is useful to note that possible roles of the
spin-phonon coupling in the peculiar electronic state of doped
cuprates have been theoretically discussed.\cite{Normand,
Jarlborg}

\section{Summary}

We measure the magnetic-field dependence of the thermal
conductivity of a parent insulating cuprate
Pr$_{1.3}$La$_{0.7}$CuO$_4$ to elucidate the possible field
dependence of phonon heat transport in cuprates. A striking dip
feature is found in $\kappa(H)$ isotherms for in-plane fields at
low temperature. The ``dip" field is found to increase linearly
with $T$ at low temperature, which is indicative of the phonon
scattering by free spins. Both the low-$T$ specific heat and the
magnetic susceptibility are measured to give supplemental evidence
for the existence of free spins. Since the Schottky-type specific
heat that signifies the existence of free spins is found
ubiquitously in high-$T_c$ cuprates in a wide doping range, the
present result suggests that the $H$-dependence of the phonon heat
transport should not be naively neglected, particularly when one
tries to extract the QP information from the quantitative analyses
of $\kappa(H)$. In addition, the present result points to a strong
spin-phonon coupling in high-$T_c$ cuprates.

\begin{acknowledgments}
We thank Y. Aoki, A. N. Lavrov, and J. Takeya for helpful
discussions. This work was supported by the Grant-in-Aid for
Science provided by the Japan Society for the Promotion of
Science.
\end{acknowledgments}

\end{document}